\documentclass[journal=achemso-jceaax,manuscript=article]{achemso}
\usepackage{natmove,rotating}

\author{Caroline Desgranges}
\author{Jerome Delhommelle}
\email{jerome.delhommelle@und.edu}
\affiliation[University of North Dakota]
{Department of Chemistry, University of North Dakota, Grand Forks, ND 58202, USA\\
(Phone: 701-777-2495)}

\title{Benchmark free energies and entropies for saturated and compressed water} 

\begin{document}

\begin{abstract}
We use molecular simulation to compute the thermodynamic properties of $7$ rigid models for water (SPC/E, TIP3P, TIP4P, TIP4P/2005, TIP4P/Ew, TIP5P, OPC) over a wide range of temperature and pressure. Carrying out Expanded Wang-Landau simulations, we obtain a high accuracy estimate for the grand-canonical partition function which, in turn, provides access to all properties, including the free energy and entropy, both at the vapor-liquid coexistence and for compressed water. The results at coexistence highlight the close connection between the behavior of the statistical functions and the thermodynamic properties. They show that the subgroup \{SPC/E,TIP4P/2005,TIP4P/Ew\} provides the most accurate account of the vapor-liquid equilibrium properties. For compressed water, the comparison of the simulation results to the experimental data establishes that the TIP4P/Ew model performs best among the $7$ models considered here, and captures the experimental trends for the dependence of entropy and molar Gibbs free energy on pressure.
\end{abstract}

\maketitle

\section{Introduction}
Water is arguably the most important molecular fluid on Earth, and is ubiquitous in biological processes and geochemical processes. Its unusual properties and behavior have long fascinated scientists and motivated a great number of research studies. In particular, its thermodynamics reveal many unexpected properties~\cite{koga2000first,stokely2010effect}, unique to water that have been connected to the existence of hydrogen bond networks in this fluid and to the complex interplay between structure and density~\cite{errington2001relationship,hummer2001water,vega2011simulating}. Accordingly, a large number of molecular models have been developed to model the properties of water. Among these, rigid water models, with a fixed charge distribution, have become widely popular and are the most commonly used~\cite{berendsen1987missing,jorgensen1983comparison,abascal2005general,horn2004development,mahoney2000five,izadi2014building}. While the relative performance of such force fields to model the pressure-density relation and the densities at the vapor-liquid coexistence has been assessed~\cite{Boulougouris,errington1998molecular,chen2000novel,stubbs2001monte,lisal2002examination,lisal2004vapor,vega2006vapor,desgranges2016ideality,desgranges2017ginzburg}, their ability to predict accurately entropy and free energy still needs to be further investigated. These are key properties in developing a full understanding of, for instance, phase changes and nucleation processes~\cite{matsumoto2002molecular,chen2005simulating,desgranges2016free}, for which the interplay between order, density and thermodynamics is expected to play a major role.

In this work, we use the Expanded Wang-Landau (EWL) simulation method~\cite{PartI,PartII,PartIII,PartIV,PartV} developed in our group to determine the free energy and entropy for $7$ rigid models for water (SPC/E~\cite{berendsen1987missing}, TIP3P~\cite{jorgensen1983comparison}, TIP4P~\cite{jorgensen1983comparison}, TIP4P/2005~\cite{abascal2005general}, TIP4P/Ew~\cite{horn2004development}, TIP5P~\cite{mahoney2000five}, OPC~\cite{izadi2014building}) over a wide range of thermodynamic conditions. EWL simulations are carried out within the grand-canonical ensemble and provide a high accuracy estimate for the value taken by the partition function. Using the statistical mechanics formalism, the knowledge of the partition function gives access to all thermodynamic properties of the system, both at the vapor-liquid coexistence and for bulk phases. In particular, here, we focus on assessing the ability of the models to correctly account for the free energy and entropy of water at the vapor-liquid coexistence, as well as for compressed water. The $7$ rigid models tested for water in this work, consist of a distribution of point charges, modeling the dipolar nature of water and of a Lennard-Jones site corresponding to the van der Waals interactions between water molecules. These models differ in a number of ways, from the set of reference properties and state points considered when parametrizing the model, to the way electrostatic interactions are accounted for. For the latter, considerations of different nature are factored in, including e.g. the computational cost (fewer interaction sites/distances calculated, e.g. $3$ atomic charges for the SPC/E and TIP3P models) or the accuracy with which electrostatic multipoles are modeled (looking beyond dipoles, i.e. at quadrupoles and higher moments, as for the OPC model). Here, we carry out a consistent evaluation of the performance of these models for the prediction of entropy and molar Gibbs free energy over a wide range of thermodynamic conditions, and establish a ranking of these rigid water models for such applications.

The paper is organized as follows. In the next section, we present the simulation method as well as the technical details and the various models studied in this work. We then present the results obtained from the EWL simulations, validate the accuracy of the results against high accuracy results obtained in recent work using flat histogram methods, and assess the ability of the models to predict reliably the molar Gibbs free energy and entropy of water phases against the experimental data at the vapor-liquid coexistence and for compressed water for pressures up to $50$~MPa. We finally draw the main conclusions from this work in the last section by establishing a ranking of the $7$ water models. 

\section{Simulation methods}
Several accurate methods for determining the free energy of molecular systems have been devised in recent years~\cite{kofke2005free}. Similarly, the determination of the conditions of phase coexistence has undergone tremendous advances with the development of the Gibbs ensemble Monte Carlo method~\cite{panagiotopoulos1987direct}, Gibbs-Duhem integration~\cite{kofke1993direct}, histogram reweighting~\cite{errington1998fixed} and flat histogram methods~\cite{Shell,errington2003direct,Camp,WLHMC,Tsvetan1,PartI,Rane1}. In this work, we are interested in determining at the same time different thermodynamic properties of the system over a wide range of conditions in a single simulation run. For this purpose, we use the EWL simulation method~\cite{PartI,PartII,PartIII,PartIV,PartV} that combines the advantages of a flat histogram sampling scheme~\cite{Wang1,Wang2,Yan,Shell} with those of the expanded ensemble approach~\cite{Paul,expanded,Lyubartsev,Shi,MV2}. It relies on a thorough sampling of all possible number of molecules of water $N$ in a given volume $V$ and at a fixed temperature $T$. This is achieved through a Wang-Landau sampling~\cite{Wang1,Wang2,Yan,Shell} of the grand-canonical ensemble $\mu VT$, with the efficiency of the sampling of all possible $N$~\cite{PartI,Mercury,Andrew} being bolstered by the expanded ensemble method~\cite{Paul,expanded,Lyubartsev,Shi,MV2}. In the expanded ensemble approach, the insertion or deletion of entire molecules is split into a finite number of stages, resulting in much greater acceptance rates for the corresponding Monte Carlo moves. The output of the EWL simulations are the grand-canonical partition function $\Theta(\mu,V,T)$ defined as
\begin{equation}
\Theta(\mu,V,T)= \sum_{N=0}^\infty Q(N,V,T) \exp (\beta \mu N)
\end{equation}
where the $Q(N,V,T)$ functions are used as biasing functions to achieve a flat histogram sampling. Once $\Theta(\mu,V,T)$ is known, we can evaluate the number distribution $p(N)$ as 
\begin{equation}
p(N) = {Q(N,V,T)  \exp (\beta \mu N)  \over \Theta(\mu,V,T)}\\
\label{distrib}   
\end{equation}

The thermodynamic properties of single phase systems (e.g. compressed water) are then obtained according to the following equations:
\begin{equation}
\begin{array}{c}
<\rho> = {\sum_{N} {N \over V} Q(N,V,T)  \exp\left(\beta \mu N)\right) \over \sum_{N} Q(N,V,T)  \exp\left(\beta \mu N)\right)}\\
U={\sum_{N} \left(E_{pot}(N)+ 3 k_B T \right) p(N) \over \sum_{N} p(N)}\\
S={k_B  \ln \Theta(\mu,V,T) \over <N>} + {\left(U - \mu \right) \over T}\\
G=U+ k_B  T \ln \Theta(\mu,V,T) -TS\\
\end{array}
\label{singlephase}   
\end{equation}
To determine the conditions for coexistence, we first vary the molar Gibbs free energy (or chemical potential $\mu$) to obtain equal probabilities for the two coexisting phases $\Pi_v=\Pi_l$ with the probability for the vapor being defined as $\Pi_v=\sum_0^{N_b} N p(N)$ and $\Pi_l=\sum_{N_b}^{N_{max}} N p(N)$ ($N_b$ is the number of molecules for which $p(N)$ reaches a minimum and $N_{max}$ is the maximum number of molecules sampled during the EWL simulation). From there, the thermodynamic properties of the two coexisting phases can be calculated from the same equations as for single phase systems (Eq.\ref{singlephase}) by changing the bounds in the sum to the [$0$,$N_b$] interval for the vapor and [$N_b$, $N_{max}$] interval for the liquid.

The $7$ water models tested in this work rely on a common set of assumptions. First, they consider the water molecule to be rigid. Second, they model the water-water interactions through a distribution of point charges of fixed values and through a single Lennard-Jones (LJ) site per water molecule. For each model, the charge distribution is fitted to reproduce the electrostatics of water, usually in the liquid under ambient conditions. This is done effectively, i.e through a large dipole moment implicitly accounting for polarization effects. The geometric and energetic LJ parameters are fitted to match, as closely as possible, a number of selected structural and thermodynamic properties (e.g. density of liquid water and atomic distribution functions). The parameters for each $7$ rigid models studied in this work, are given in Table~\ref{Tab1}

\vspace{1cm}
\begin{table}[h]
\small
  \caption{Force field parameters for the $7$ water models studied in this work, with $q$ denoting the charge on one of the positively charged sites ($H$ atoms), $l$ is the $O-H$ bond length, $z_1$ is the offset from the $O$ atom of the negative point charge, $\theta$ is the $H-O-H$ angle and ($\sigma$, $\epsilon$) are the LJ parameters on the $O$ atom.}
  \vspace{1cm}
  \label{Tab1}
  \begin{tabular*}{\textwidth}{@{\extracolsep{\fill}}|c|c|c|c|c|c|c|c|}
\hline
$$ & $SPC/E $ &  $TIP3P $ & $TIP4P$ & $TIP4P/2005$ & $TIP4P/Ew$ & $TIP5P$ & $OPC$\\
\hline
$q (e)$ & $0.4238$ &  $0.417$ & $0.52$ & $0.5564$ & $0.5242$ & $0.241$ & $0.6791$\\
$l $~(\AA) & $1.0$ &  $0.9572$ & $0.9572$ & $0.9572$ & $0.9572$ & $0.9572$ & $0.8724$\\
$z_1$~(\AA) & $N/A^1$ &  $N/A^1$ & $0.15$ & $0.1546$ & $0.125$ & $N/A^2$ & $0.1594$\\
$\theta (deg)$ & $109.47$ &  $104.52$ & $104.52$ & $104.52$ & $104.52$ & $104.52$ & $103.6$\\
$\sigma_{LJ}$~(\AA) & $3.166$ &  $3.15061$ & $3.15365$ & $3.1589$ & $3.16435$ & $3.12$ & $3.16655$\\
$\epsilon_{LJ} (kJ/mol)$ & $0.65$ &  $0.6364$ & $0.6480$ & $0.7749$ & $0.680946$ & $0.6694$ & $0.89036$\\
\hline
\end{tabular*}
\end{table}

EWL simulations are run using the same parameters as discussed in previous work~\cite{PartIII,desgranges2016ideality}. Specifically, the number of stages is set to $100$ for the expanded ensemble approach, the starting value of the convergence factor $f$ is set to $e$ and the final value of $f$ is fixed to $10^{-8}$, with a minimum number of visits of $1000$ per interval for the number of molecules. EWL simulations are carried out within a Monte Carlo framework. $25\%$ of the attempted MC moves are translation of a single molecule, $25\%$ are rotation of a single molecule and the rest of the attempted MC moves are changes in the number of molecules (insertion/deletion). Simulation results are obtained by running a EWL simulation at a given temperature between $300$~K and $475$~K with $25$~K interval and in a cubic cell of a $20$~\AA~edge, with the usual periodic boundary conditions applied (at low temperature, additional simulations with a larger volume are run to obtain an accurate value for the density of the vapor phase). The LJ interactions are truncated at half the boxlength $L$, and the usual tail correction are applied beyond. We also use the Ewald sum technique to calculate the long-range interactions between water molecules with a screening parameter for the charge gaussian distribution set to $5.6/L$ and a reciprocal cutoff vector set to $k_{max}=6 \times (2 \pi/L)$. 
Throughout the EWL simulations, we collect histograms for the number of times each $N$ value is visited to ensure that all $N$ are sampled a large number of times, as well as for the potential energy of the system associated with each $N$ value to allow then for the calculation of e.g. the internal energy and enthalpy of the system. During the EWL simulations, a third histogram (for the biasing function $\ln Q(N,V,T)$) is dynamically updated each time a given value for $N$ is reached. This histogram is later used to calculate the grand-canonical partition function and all thermodynamic properties through the statistical mechanics formalism and Eq.~\ref{singlephase}. The computational cost of the EWL method is similar to that involved in other flat histogram methods, such as e.g. the transition matrix Monte Carlo (TMMC) method~\cite{errington2003direct}, and yields results in good agreement with these methods, as discussed in the next section. Error estimates for the simulation results are evaluated as follows. We carry out $4$ independent EWL simulations starting from $N=0$ and using different seeds for the random number generator. This provides a set of numerical data over which we can calculate standard deviations, both for the partition functions and for the thermodynamic functions through Eq.~\ref{singlephase}.

\section{Results and discussion}

\begin{figure}
\begin{center}
\includegraphics*[width=8cm]{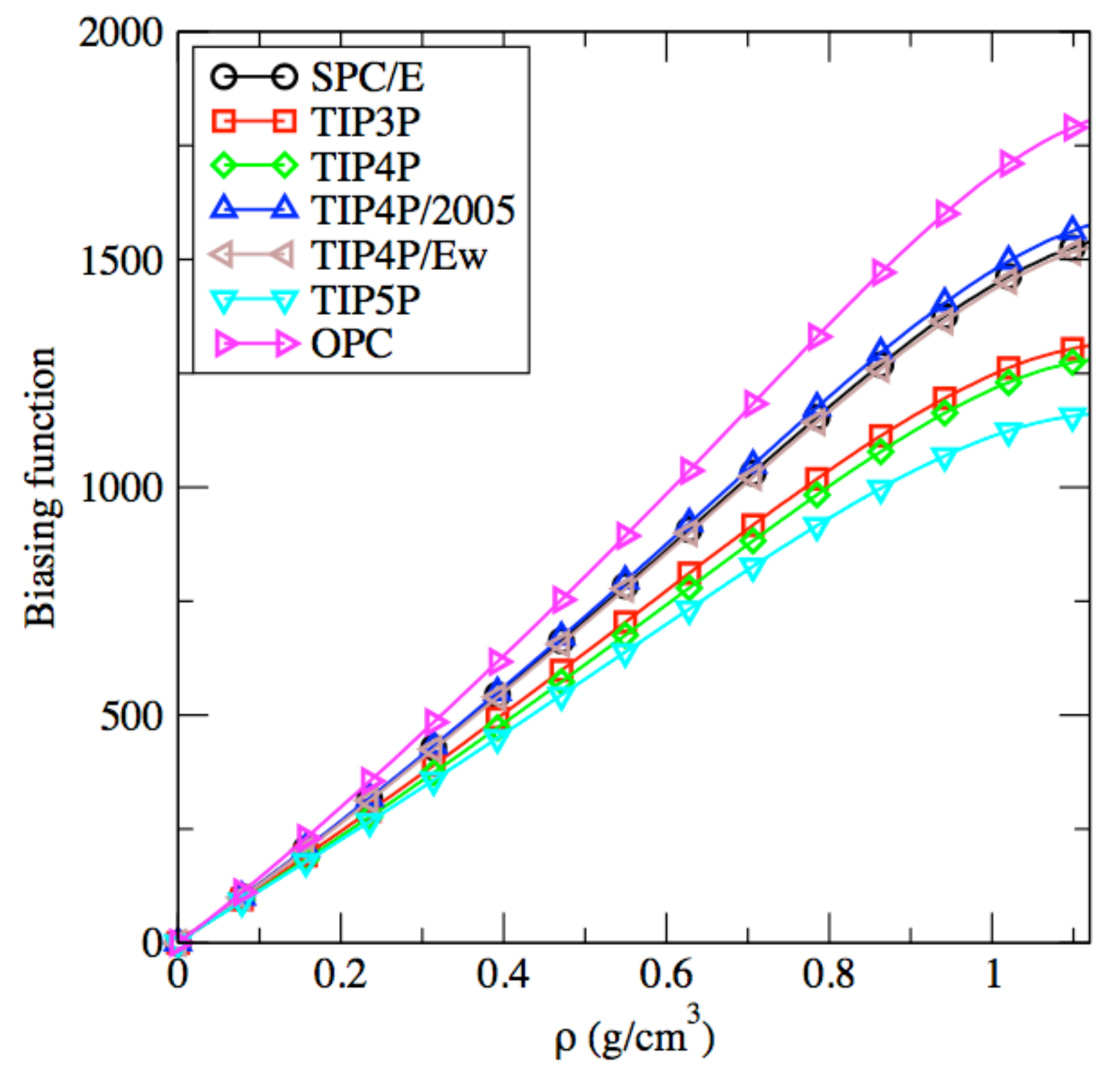}
\end{center}
\caption{Biasing function for the EWL simulations (error bars are smaller than the symbol size). Results are shown for $T=475$~K for all rigid models studied in this work.}
\label{Fig1}
\end{figure}

\begin{figure}
\begin{center}
\includegraphics*[width=8cm]{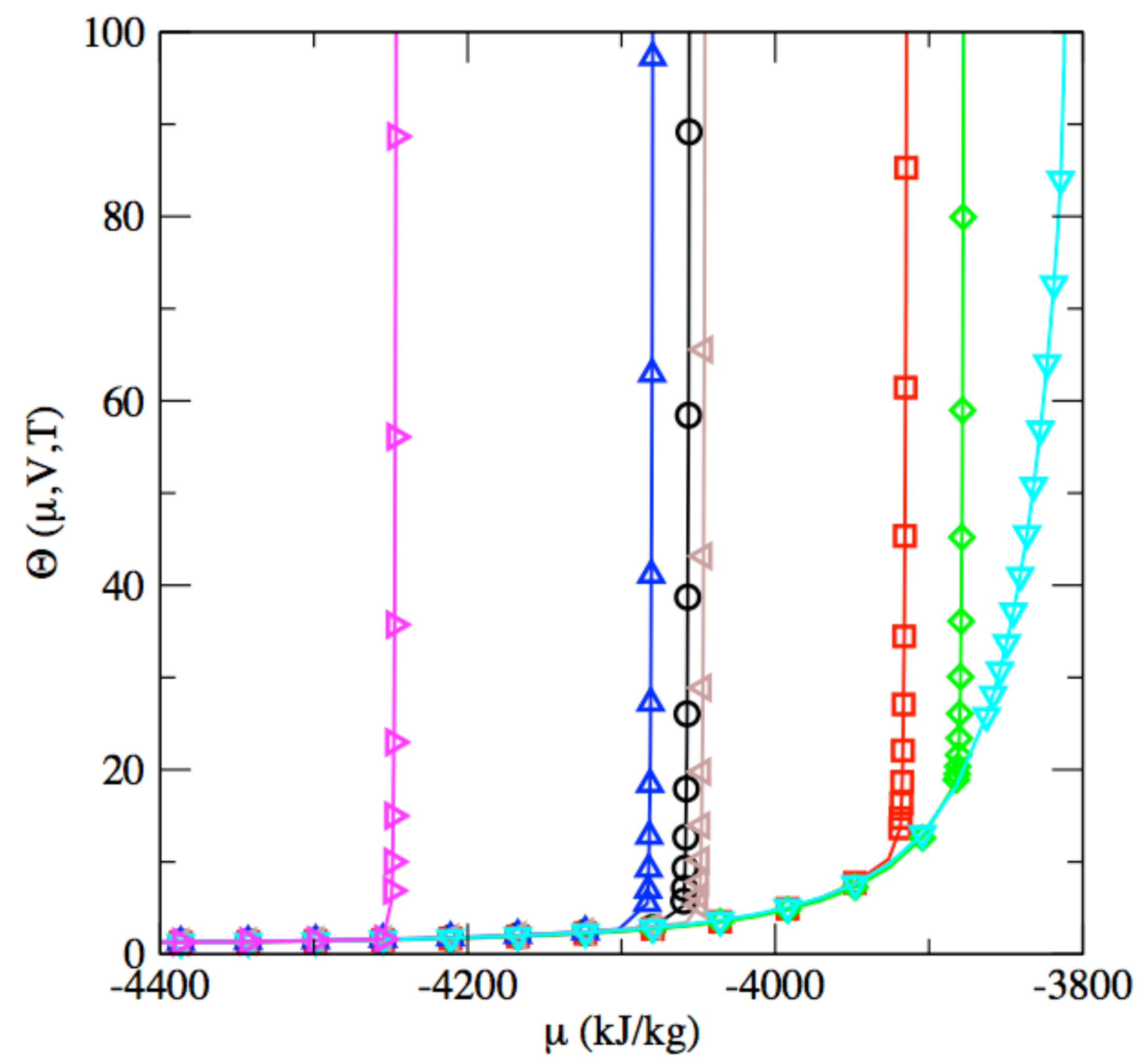}
\end{center}
\caption{Grand-canonical partition function at $T=475$~K (error bars are smaller than the symbol size). For each model, the sharp change in the function behavior indicates the vapor $\to$ liquid transition (same legend as in Fig~\ref{Fig1}).}
\label{Fig2}
\end{figure}

We show in Fig.~\ref{Fig1} and Fig.~\ref{Fig2}, the output obtained from the EWL simulations for the different water models at $T=475$~K. Starting with the results for the biasing functions (Fig.~\ref{Fig1}), we find that the type of force field used, and the underlying assumptions made during its parametrization, have a significant impact on the statistical functions associated with water. Comparing the results obtained for, e.g., a density of $1$~$g/cm^3$, we find that the value taken by the biasing function for the OPC model is $34$~\% greater than that of the TIP5P model. In between, we find results that can be gathered into broadly two groups. First, on the lower end of the spectrum, about $10$~\% above the TIP5P value, we have two force fields, composing the \{TIP3P, TIP4P\} subgroup. Second, at about $23$~\% above the TIP5P model, we find another subgroup, including the three other force fields, i.e. the \{TIP4P/2005, TIP4P/Ew and SPC/E\} subgroup, for which very similar values for the biasing function are obtained. 

We now discuss how the different values obtained for the biasing function are connected to the ability of the models to predict reliably the thermodynamic properties of water. We start by examining the second statistical function, $\Theta (\mu, V,T)$ and analyze the impact of the choice of a specific model on the partition function. We observe in Fig.~\ref{Fig2} that the partition function exhibits a sharp increase beyond a threshold value for $\mu$, as shown e.g., in the case of the OPC model, for$\mu$ greater than $\mu_{OPC}=-4260$~kJ/kg. Most notably, we find that the order in which this steep increase occurs for the different models mirrors that for the biasing functions of Fig~\ref{Fig1}. Specifically, we have: $\mu_{OPC} <  \mu_{TIP4P/2005} < \mu_{SPC/E} < \mu_{TIP4P/Ew}< \mu_{TIP3P}< \mu_{TIP4P}< \mu_{TIP5P}$. We also see that the results for the OPC and the TIP5P models define the lower and upper bounds, respectively, for the range of results obtained for the other models, with again the same $2$ subgroups appearing clearly in line with our findings for the biasing functions. There, the connection between the statistical mechanics functions and the thermodynamic behavior is starting to emerge, since the sharp increase in the partition function is known to be related to the onset of the vapor $\to$ liquid transition~\cite{jctc2015}. Furthermore, we find that the value of $\Theta (\mu, V, T)$ at the threshold value is model-dependent with e.g. $\Theta (\mu_{OPC}, V, T)$ < $\Theta (\mu_{TIP5P}, V, T)$. This has a direct consequence on the thermodynamics at coexistence. Since the saturation pressure $P_{sat}$ is proportional to the logarithm of $\Theta(\mu,V,T)$ (see Eq.~\ref{singlephase}), this means that the OPC model will predict a lower saturation pressure while TIP5P will yield the greater value for $P_{sat}$.

\begin{figure}
\begin{center}
\includegraphics*[width=8cm]{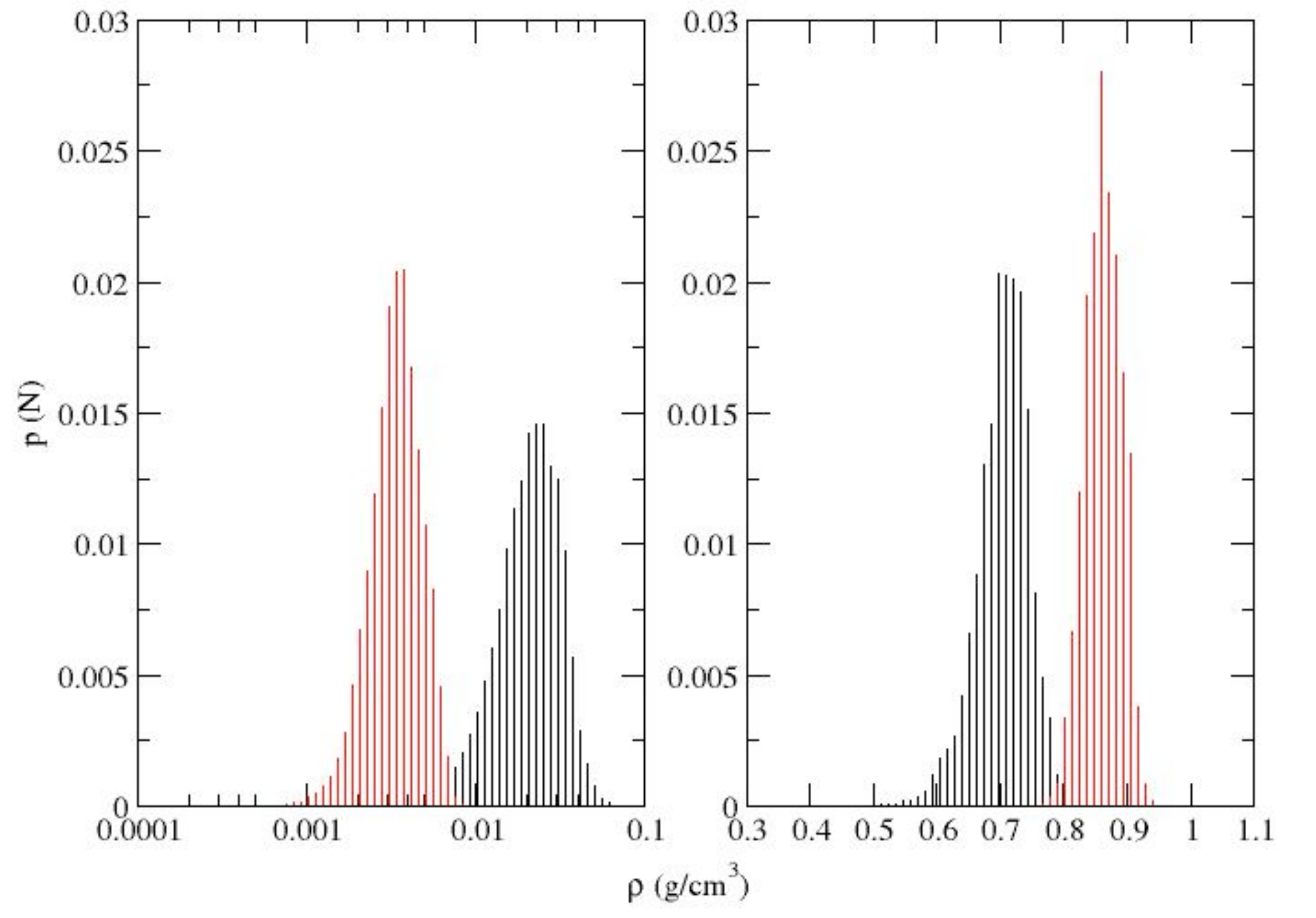}
\end{center}
\caption{Number distribution, as a function of density, for the TIP5P model at ($T=400$~K, $\mu=-3373.53$~kJ/kg) in red, and at ($T=475$~K, $\mu=-3809.21$~kJ/kg) in black. Results are shown for the vapor on a logarithmic scale (left panel) and on a linear scale for the liquid (right panel)}
\label{Fig3}
\end{figure}

Before carrying out a detailed analysis of the performance of the $7$ models, we discuss how the densities at coexistence are determined. Starting from the results for the biasing functions, we systematically vary the value of $\mu$ and calculate the number distribution $p(N)$ for each $\mu$. Once we find the value of $\mu$ that leads to two peaks of equal area for $p(N)$ or, equivalently, to equal probabilities for the vapor and for the liquid phase, we obtain the following plots shown in Fig~\ref{Fig3} on the example of the TIP5P model, as shown here at $T=400$~K ($\mu=-3373.53$~kJ/kg) and at $T=475$~K ($\mu=-3809.21$~kJ/kg). This distribution provides a direct link to the densities at coexistence, with the left peak corresponding to the vapor at coexistence and the right peak to the liquid at coexistence. It also shows how the binodal closes as temperature increases, with the two peaks at $T=475$~K being closer to each other than at $T=400$~K. Repeating this process for temperatures between $300$~K and $475$~K, with a $25$~K increment, we determine the vapor-liquid equilibria of water for the $7$ rigid models studied in this work.

To validate our results against simulation data from the literature, we compare in Table~\ref{Tab2} the results obtained for the SPC/E model using the EWL method to high-accuracy simulation data obtained in previous work with another flat histogram method, the Transition Matrix Monte Carlo method (TMMC)~\cite{errington2003direct,NIST}. We summarize in Table~\ref{Tab2} the results for the densities at coexistence, as well as for $P_{sat}$. As shown in Table~\ref{Tab2}, there is overall a good agreement between the EWL and TMMC results, with e.g. a standard deviation of $4.4 \times 10^{-3}$~g/cm${^3}$ for the liquid densities. Similarly, the two methods yield vapor densities and saturation pressures that are within the statistical uncertainty for most of the temperature range, the agreement becoming lease satisfactory at high pressure as the system gets closer to the critical point. Using a scaling law to determine the critical temperature and the law of rectilinear diameters to evaluate the critical density for the SPC/E model, we check that we obtain a critical point ($T_c^{EWL}=641\pm5$~K and $\rho_c^{EWL}=0.31\pm0.01$~g/cm$^3$) in good agreement with prior work using the Gibbs Ensemble Monte Carlo (GEMC) method~\cite{Boulougouris}. We add that, when compared to the GEMC method, the advantage of the EWL method, as well as of other methods implemented in the grand-canonical ensemble. like the histogram reweighting method~\cite{potoff1998critical} and the TMMC method~\cite{errington2003direct}), is that a finite-size scaling approach can be applied to determine accurately the critical point. This is now stated in the revised manuscriptWe also check that the EWL results are consistent with, when available, the literature data on other models, and using other methods, such as e.g. the Gibbs-Duhem integration method for the TIP4P/2005 and TIP4P/Ew~\cite{vega2006vapor} and the Gibbs Ensemble Monte Carlo method for the TIP5P model~\cite{lisal2002examination,lisal2004vapor}. Results for this comparison are provided in Table~\ref{Tab3} for the VLE properties at 300~K, and show a good agreement between the results obtained with the EWL methods and other simulation methods.

\begin{table}[h]
\small
  \caption{Comparison between the EWL (this work) and the TMMC results~\cite{errington2003direct,NIST} for the phase equilibria of the SPC/E model for water. $\rho$~is given in $g/cm^3$~and $T$ in $K$ (errors bars are of the order of $1$~\% for $\rho_v$, $0.5$~\% for $\rho_l$ and $1.5$~\% for $P_{sat}$)}
  \label{Tab2}
  \begin{tabular*}{\textwidth}{@{\extracolsep{\fill}}|c|c|c|c|c|c|c|}
\hline
$T$ & $\rho_v^{EWL} $ &  $\rho_v^{TMMC} $ & $\rho_l^{EWL} $ & $\rho_v^{TMMC} $ & $P_{sat}^{EWL} $ & $P_{sat}^{TMMC} $\\
\hline
$300$ & $7.38\times10^{-6}$ &  $7.55\times10^{-6}$ & $1.000$ & $0.99692$ & $0.011$& $0.0104$ \\
$325$ & $3.14\times10^{-5}$ &  $3.099\times10^{-5}$ & $0.981$ & $0.982$ & $0.049$& $0.04605$ \\
$350$ & $1.02\times10^{-4}$ &  $1.026\times10^{-4}$ & $0.959$ & $0.964$ & $0.169$& $0.1628$ \\
$375$ & $2.79\times10^{-4}$ &  $2.797\times10^{-4}$ & $0.938$ & $0.9423$ & $0.502$& $0.469$ \\
$400$ & $6.64\times10^{-4}$ &  $6.68\times10^{-4}$ & $0.920$ & $0.920$ & $1.287$& $1.171$ \\
$425$ & $1.46\times10^{-3}$ &  $1.429\times10^{-3}$ & $0.886$ & $0.8918$ & $2.799$& $2.59$ \\
$450$ & $2.87\times10^{-3}$ &  $2.82\times10^{-3}$ & $0.856$ & $0.863$ & $5.659$& $5.20$ \\
$475$ & $5.06\times10^{-3}$ &  $5.21\times10^{-3}$ & $0.825$ & $0.8293$ & $9.950$& $9.65$ \\
\hline
\end{tabular*}
\end{table}

\begin{table}[h]
\small
  \caption{Prediction of water VLE properties at $300$~K: comparison between the results obtained for different models using either the EWL method (this work) or other simulation methods (SIM), i.e. the TMMC results for the TIP3P and TIP4P models~\cite{errington2003direct,NIST}, the Gibbs-Duhem simulation method for the TIP4P/2005 and TIP4P/Ew models~\cite{vega2006vapor} and the Gibbs Ensemble Monte Carlo method for the TIP5P model~\cite{lisal2004vapor} (SIM data were obtained at $298.15$~K for the TIP5P model). Same units and errors bars as in Table 2.}
  \label{Tab3}
  \begin{tabular*}{\textwidth}{@{\extracolsep{\fill}}|c|c|c|c|c|c|c|}
\hline
Model & $\rho_v^{EWL} $ &  $\rho_v^{SIM} $ & $\rho_l^{EWL} $ & $\rho_v^{SIM} $ & $P_{sat}^{EWL} $ & $P_{sat}^{SIM} $\\
\hline
TIP3P & $3.95\times10^{-5}$ &  $3.74\times10^{-5}$ & $0.978$ & $0.9839$ & $0.052$& $0.051$ \\
TIP4P & $3.76\times10^{-5}$ &  $3.739\times10^{-5}$ & $0.990$ & $0.9925$ & $0.056$& $0.05137$ \\
TIP4P/2005 & $5.67\times10^{-6}$ &  $5.642\times10^{-6}$ & $0.998$ & $0.9965$ & $0.008$& $0.00778$ \\
TIP4P/EW & $8.41\times10^{-6}$ &  $8.201\times10^{-6}$ & $0.995$ & $0.994$ & $0.012$& $0.013$ \\
TIP5P & $7.79\times10^{-5}$ &  $8.5\times10^{-5}$ & $0.990$ & $0.984$ & $0.110$& $0.116$ \\
\hline
\end{tabular*}
\end{table}

\begin{figure}
\begin{center}
\includegraphics*[width=10cm]{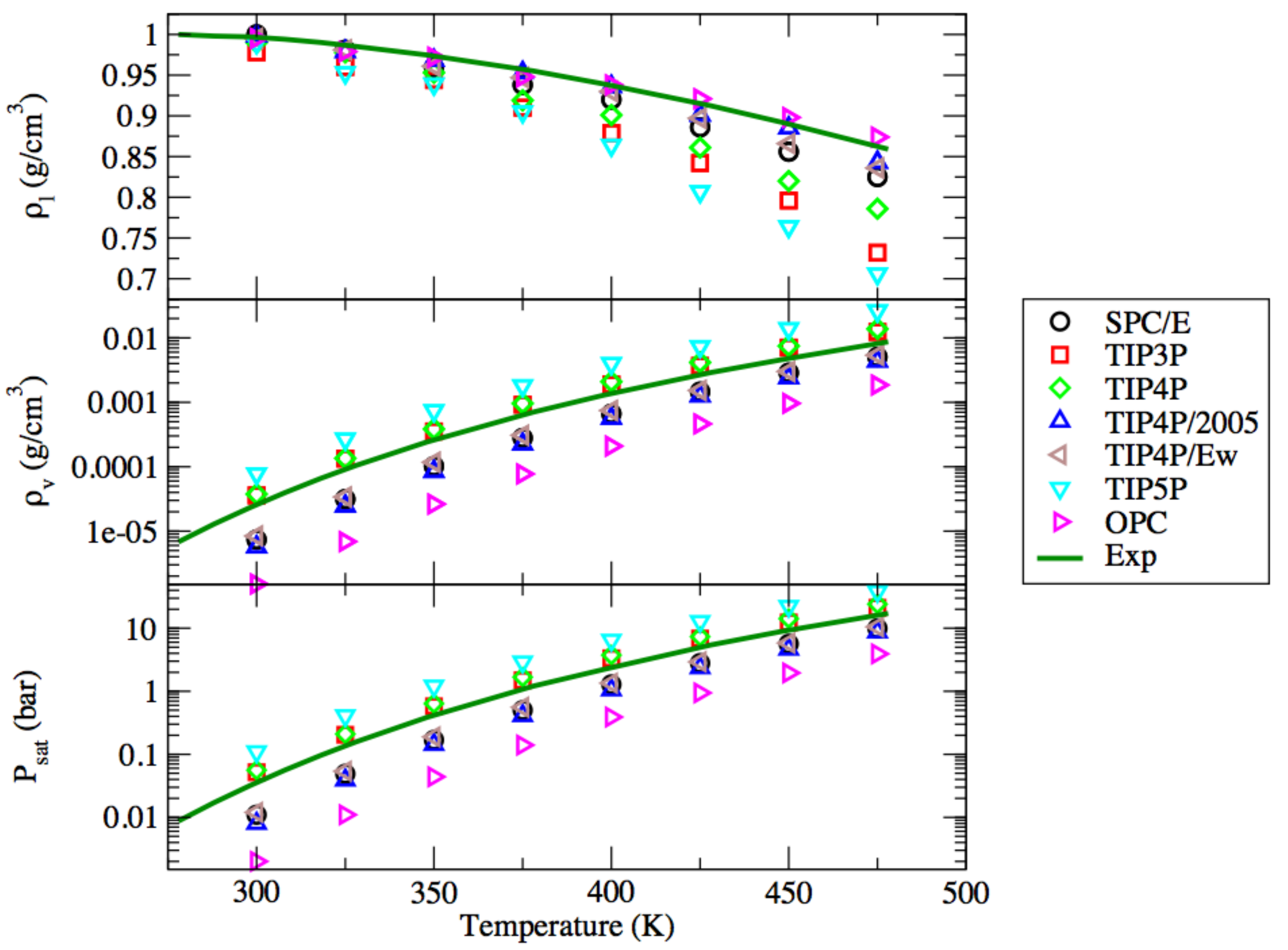}
\end{center}
\caption{Vapor-liquid equilibria of water for the rigid models used in this work. (Top) Liquid densities, (Middle) Vapor densities and (Bottom) Saturation pressure. Experimental data are taken from the IAPWS tables~\cite{wagner2002iapws}.}
\label{Fig4}
\end{figure}

We now discuss the relative performance of the $7$ models by comparing the predicted values for the saturation properties to the experimental data~\cite{wagner2002iapws}. We show in Fig.~\ref{Fig4}  the liquid densities in the top panel, the vapor densities (plotted with a logarithmic scale) in the middle panel and the saturation pressures in the bottom panel. We observe that there is a strong correlation between the results obtained for the statistical functions (biasing and partition functions) and the results obtained for the densities at coexistence. Specifically, the densities for the OPC and TIP5P model define the bounds for intervals spanned by the results obtained for the other models, with the OPC model yielding the lowest vapor densities, the lowest saturation pressures  and the largest liquid densities of the $7$ models. On the other hand, the TIP5P model gives the largest vapor densities, the largest saturation pressures and the lowest liquid densities, with the results for the other models being organized in the same two subgroups, i.e. \{TIP4P/2005, TIP4P/Ew,SPC/E\}  and \{TIP3P, TIP4P\}, as in Figs.~\ref{Fig1} and~\ref{Fig2}. 

\begin{sidewaystable}[!ph]
\caption{Deviation from the experimental data~\cite{wagner2002iapws} of the predicted $\rho_{vap}$, $\rho_{liq}$ and $P_{sat}$ for the $7$ models considered in this work (error bars are are of the order of $1$~\% for $\rho_v$, $0.5$~\% for $\rho_l$ and $1.5$~\% for $P_{sat}$)}
\label{Tab4}
\centering
\begin{tabular}{|c|c|c|c|c|c|c|c|c|}
\hline
$T$ & $\rho_v^{Exp} $ &  $SPC/E$ & $TIP3P$ & $TIP4P$ & $TIP4P/2005$ & $TIP4P/Ew$ & $TIP5P$ & $OPC$\\
\hline
$300$ & $2.55\times10^{-5}$ & $-1.812\times10^{-5}$ & $+1.04\times10^{-5}$ & $+1.21\times10^{-5}$ & $-1.983\times10^{-5}$ & $-1.709\times10^{-5}$& $+5.24\times10^{-5}$ & $-2.399\times10^{-5}$ \\
$325$ & $9.06\times10^{-5}$ & $-5.92\times10^{-5}$ & $+4.34\times10^{-5}$ & $+4.44\times10^{-5}$ & $-6.61\times10^{-5}$ & 
$-5.67\times10^{-5}$& $+1.784\times10^{-4}$ & $-8.37\times10^{-5}$\\
$350$ & $2.60\times10^{-4}$ & $-1.58\times10^{-4}$ & $+9.4\times10^{-5}$ & $+1.25\times10^{-4}$ & $-1.758\times10^{-4}$ & 
$-1.43\times10^{-4}$& $+4.81\times10^{-4}$ & $-2.337\times10^{-4}$\\
$375$ & $6.37\times10^{-4}$ & $+3.58\times10^{-4}$ & $+2.82\times10^{-4}$ & $+3.24\times10^{-4}$ & $-4.13\times10^{-4}$ & 
$-3.29\times10^{-4}$& $+1.173\times10^{-3}$ & $-5.597\times10^{-4}$\\
$400$ & $1.37\times10^{-3}$ & $-7.06\times10^{-4}$ & $+5.2\times10^{-4}$ & $+7.2\times10^{-4}$ & $-8.05\times10^{-4}$ & 
$-6.25\times10^{-4}$& $+2.62\times10^{-3}$ & $-1.161\times10^{-3}$\\
$425$ & $2.67\times10^{-3}$ & $-1.21\times10^{-3}$ & $+1\times10^{-3}$ & $+1.49\times10^{-3}$ & $-1.41\times10^{-3}$ & 
$-1.14\times10^{-3}$& $+4.65\times10^{-3}$ & $-2.208\times10^{-3}$ \\
$450$ & $4.82\times10^{-3}$ & $-1.95\times10^{-3}$ & $+2.26\times10^{-3}$ & $+2.66\times10^{-3}$ & $-2.41\times10^{-3}$ & 
$-1.82\times10^{-3}$& $+9.08\times10^{-3}$ & $-3.856\times10^{-3}$ \\
$475$ & $8.17\times10^{-3}$ & $+2.4\times10^{-4}$ & $+4.23\times10^{-3}$ & $+5.63\times10^{-3}$ & $-3.8\times10^{-3}$ & 
$-2.78\times10^{-3}$& $+0.01803$ & $-6.31\times10^{-3}$\\
\hline
\hline
$T$ & $\rho_l^{Exp} $ & $SPC/E$ & $TIP3P$ & $TIP4P$ & $TIP4P/2005$ & $TIP4P/Ew$ & $TIP5P$ & $OPC$\\
\hline
$300$ & $0.997$ &  $+0.003$ & $-0.019$ & $-0.007$ & $+0.001$ & $-0.002$ & $-0.007$ & $-0.002$ \\
$325$ & $0.987$ &  $-0.006$ & $-0.027$ & $-0.009$ & $-0.008$ & $-0.006$ & $-0.034$ & $-0.008$ \\
$350$ & $0.974$ &  $-0.015$ & $-0.03$ & $-0.021$ & $-0.006$ & $-0.013$ & $-0.035$ & $-0.002$ \\
$375$ & $0.969$ &  $-0.031$ & $-0.059$ & $-0.05$ & $-0.016$ & $-0.022$ & $-0.064$ & $-0.021$ \\
$400$ & $0.937$ &  $-0.017$ & $-0.058$ & $-0.036$ & $-0.001$ & $-0.007$ & $-0.073$ & $+0.001$ \\
$425$ & $0.915$ &  $-0.029$ & $-0.073$ & $-0.054$ & $-0.014$ & $-0.018$ & $-0.108$ & $+0.006$ \\
$450$ & $0.890$ &  $-0.034$ & $-0.094$ & $-0.07$ & $-0.005$ & $-0.024$ & $-0.126$ & $+0.008$ \\
$475$ & $0.862$ &  $-0.037$ & $-0.13$ & $-0.076$ & $-0.019$ & $-0.026$ & $-0.156$ & $+0.012$ \\
\hline
\hline
$T$ & $P_{sat}$ & $SPC/E$ & $TIP3P$ & $TIP4P$ & $TIP4P/2005$ & $TIP4P/Ew$ & $TIP5P$ & $OPC$\\
\hline
$300$ & $0.036$ &  $-0.025$ & $+0.016$ & $+0.02$ & $-0.028$ & $-0.024$ & $+0.074$ & $-0.034$ \\
$325$ & $0.137$ &  $-0.088$ & $+0.067$ & $+0.073$ & $-0.098$ & $-0.083$ & $+0.273$ & $-0.126$ \\
$350$ & $0.421$ &  $-0.252$ & $+0.158$ & $+0.218$ & $-0.278$ & $-0.232$ & $+0.787$ & $-0.377$ \\
$375$ & $1.092$ &  $-0.59$ & $+0.402$ & $+0.588$ & $-0.682$ & $-0.534$ & $+1.825$ & $-0.952$ \\
$400$ & $2.475$ &  $-1.188$ & $+0.868$ & $+1.27$ & $-1.423$ & $-1.133$ & $+3.983$ & $-2.085$ \\
$425$ & $5.031$ &  $-2.232$ & $+1.842$ & $+2.329$ & $-2.665$ & $-2.106$ & $+7.735$ & $-4.086$ \\
$450$ & $9.367$ &  $-3.708$ & $+2.977$ & $+4.827$ & $-4.71$ & $-3.528$ & $+12.927$ & $-7.412$ \\
$475$ & $12.227$ &  $-2.278$ & $+8.865$ & $+11.928$ & $-3.49$ & $-1.682$ & $+25.96$ & $-8.298$ \\
\hline
\end{tabular}
\end{sidewaystable}

How well do these models perform when compared to the experimental data? We provide in Table~\ref{Tab3} the absolute deviations from the reference data~\cite{wagner2002iapws} for the $7$ models. Table~\ref{Tab3} shows that both the saturation pressure and the vapor line of the reference vapor-liquid equilibrium (VLE) (see left panel of Fig.~\ref{Fig4}) fall between the two subgroups, being closely surrounded by the \{TIP4P/2005, TIP4P/Ew,SPC/E\} subgroup and by the \{TIP3P, TIP4P\} subgroup . Furthermore, the liquid part of the experimental binodal is found to be between the predictions of the OPC model, which predicts quite accurately the liquid densities, and the \{TIP4P/2005, TIP4P/Ew,SPC/E\} subgroup. On the basis of the simulation data for the densities at coexistence and the saturation pressure, the \{TIP4P/2005, TIP4P/Ew,SPC/E\} subgroup provides a better account of the VLE properties than the other models.

\begin{figure}
\begin{center}
\includegraphics*[width=6.8cm]{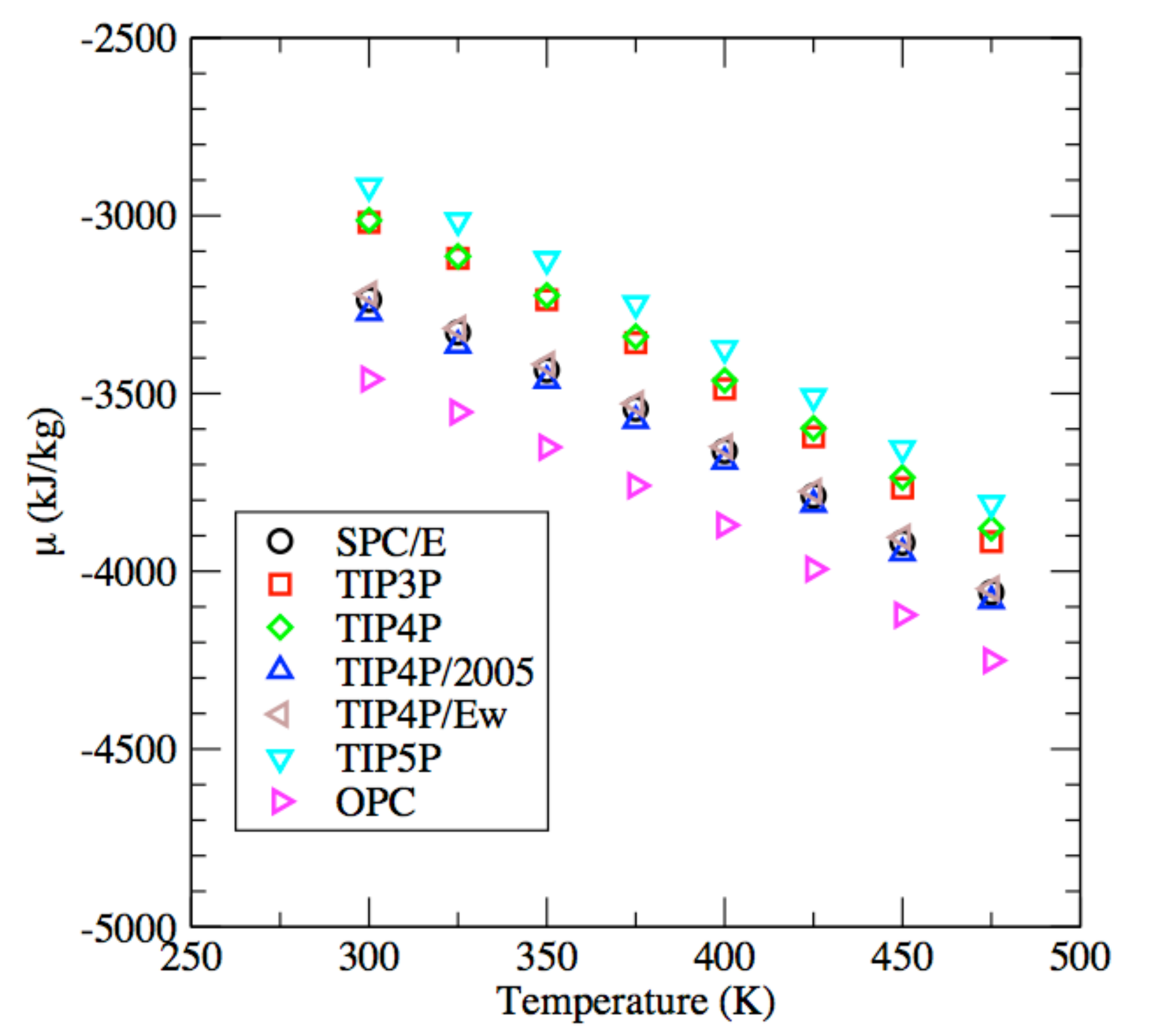}(a)
\includegraphics*[width=6.5cm]{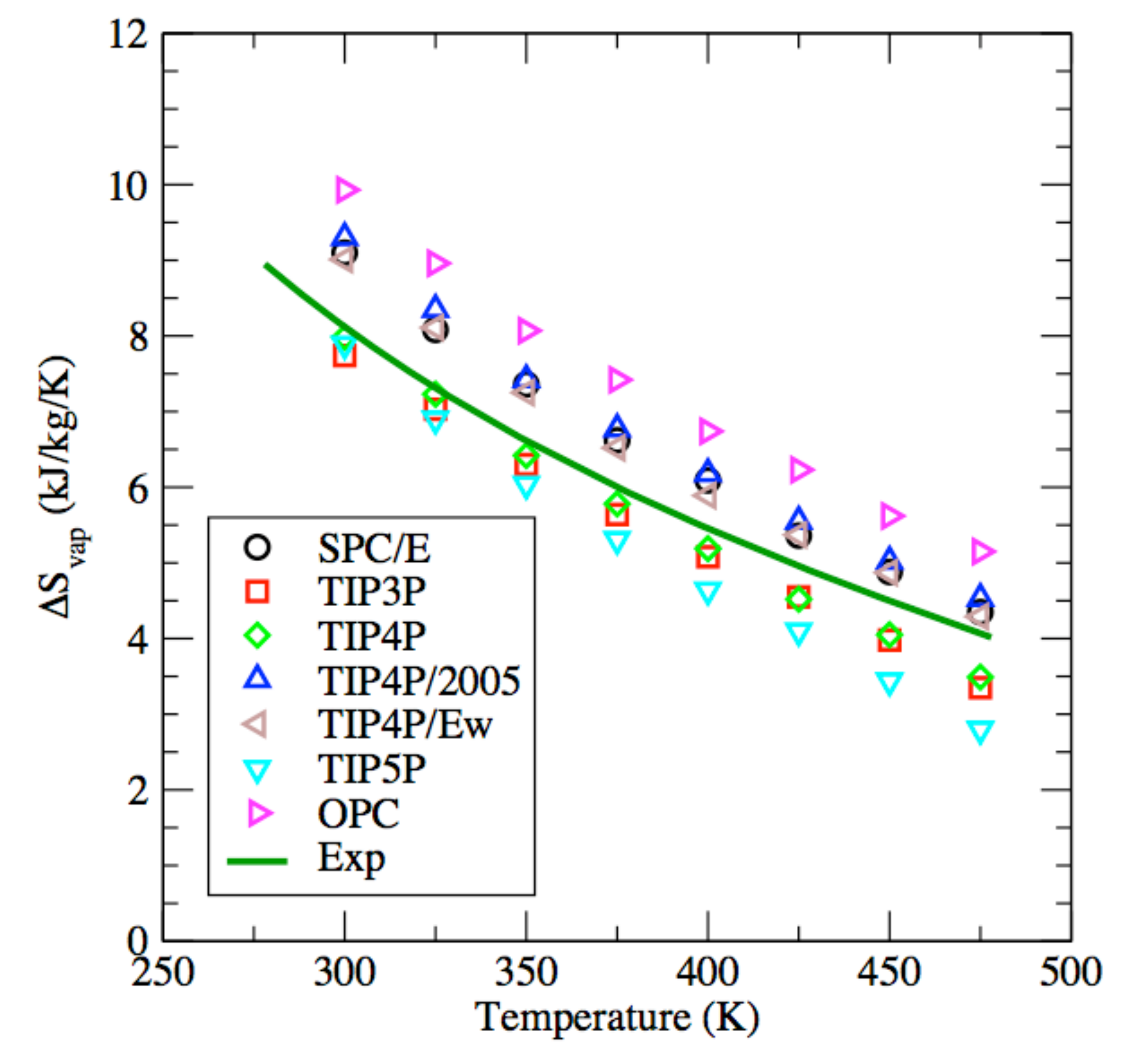}(b)
\end{center}
\caption{(a) $\mu$ at the vapor-liquid coexistence, and (b) entropy of vaporization $\Delta S_{vap}$ for the rigid models used in this work for water (error bars are smaller than the symbol size). The experimental data are shown as a green line.}
\label{Fig5}
\end{figure}

We now turn to the results for the molar Gibbs free energy $\mu$ and for the entropy of vaporization for the $7$ models studied in this work. Fig.~\ref{Fig5}(a) shows the impact of the choice of a specific rigid force field to model water on the molar Gibbs free energy at coexistence. We see that the results for the various force fields follow the same trend as previously observed, with a shift towards the greater $\mu$ as we move from the OPC model (lowest $\mu$ at coexistence) to the TIP5P model (highest $\mu$ at coexistence). We also recover the same distribution, in terms of subgroups, confirming the strong connection between the thermodynamic properties at coexistence and the statistical functions plotted in Fig.~\ref{Fig1} and Fig.~\ref{Fig2}. Moving on to the entropy of vaporization $\Delta S_{vap}$, shown in Fig.~\ref{Fig5}(b), we obtain the reverse order for the predictions of the models, with the lowest $\Delta S_{vap}$ being obtained for the TIP5P model and the highest value for the OPC model. This can be interpreted in terms of the relative values for the densities at coexistence, especially for the vapor phase. The TIP5P model predicts a very high vapor density at coexistence, and hence to a low entropy for the vapor, which, in turn, results in a much smaller $\Delta S_{vap}$. Comparing now the predictions for the models to the experimental data for $\Delta S_{vap}$, we find that the experiment falls in between the predicted values of the two subgroups, with the \{TIP3P,TIP4P\} subgroup performing best at low temperature (around $T=300$~K) and the \{TIP4P/2005,TIP4P/Ew\} subgroup yielding the most accurate results at high temperature (for $450$~K and above). This can be accounted for as follows. At low temperature, the $\rho_{vap}$ is very small and the corresponding entropy is very large. This means that an accurate prediction of the vapor density will be the most important factor in correctly accounting for $\Delta S_{vap}$, and explains why the \{TIP3P, TIP4P\} subgroup provides results that are closest to the experiment under such conditions. On the other hand, at high temperature, the density of the vapor becomes of the same order as that of the liquid, resulting in the better performance of the \{TIP4P/2005, TIP4P/Ew\} subgroup at high temperature.

\begin{figure}
\begin{center}
\includegraphics*[width=8cm]{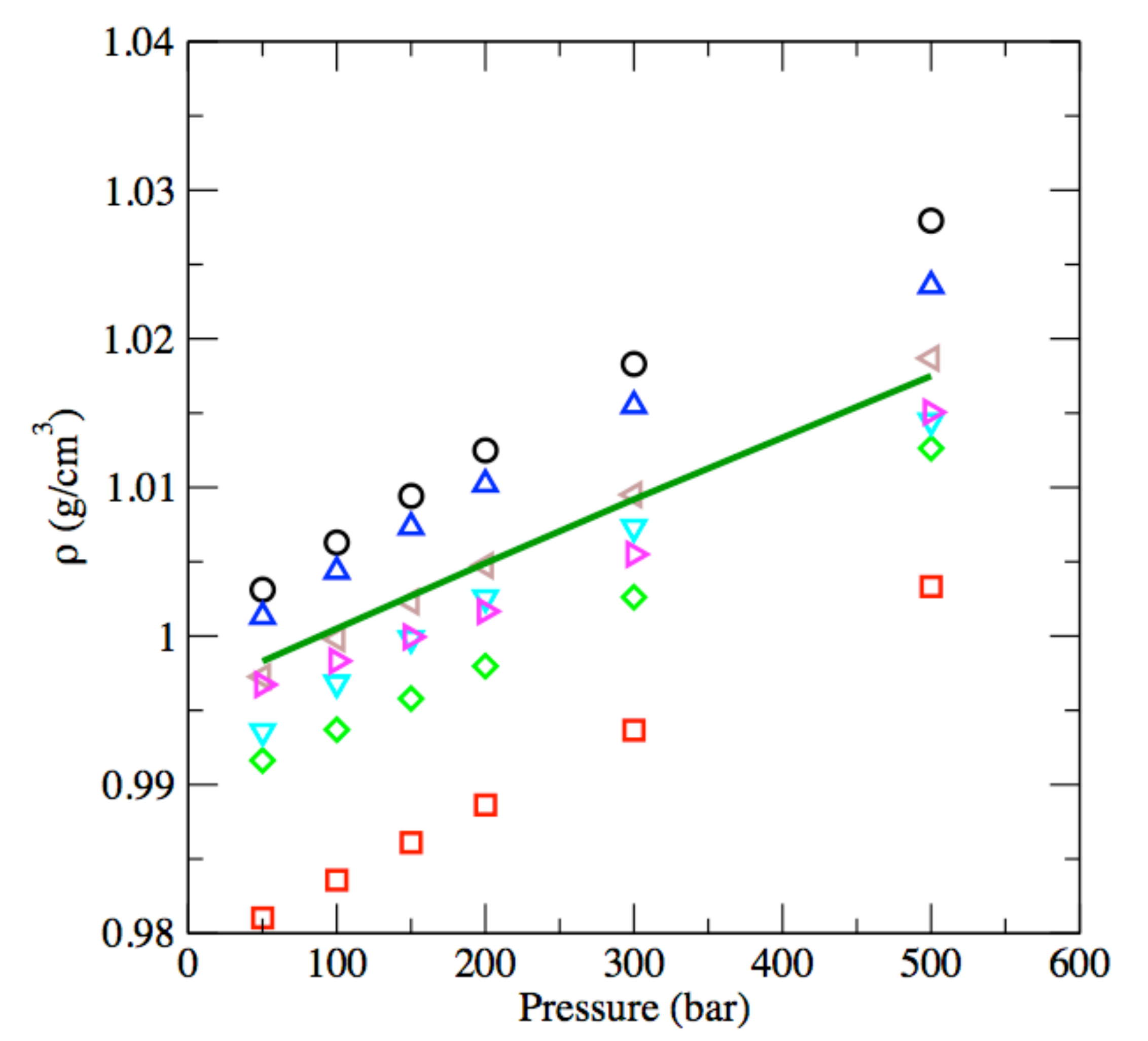}
\end{center}
\caption{Density of compressed water at $T=300$~K (same legend as in Fig.~\ref{Fig5}). Error bars are smaller than the symbol size.}
\label{Fig6}
\end{figure}

\begin{table}[ht]
\small
  \caption{Standard deviations from the experiment for compressed liquid water at 300~K.}
  \vspace{1cm}
  \label{Tab5}
  \begin{tabular*}{\textwidth}{@{\extracolsep{\fill}}|c|c|c|c|c|c|c|c|}
\hline
  &  SPC/E & TIP3P & TIP4P & TIP4P/2005 & TIP4P/Ew & TIP5P & OPC \\
\hline
$\sigma_{\rho}$ $(g/cm^3)$ & $0.0076$ &  $0.0162$ & $0.0065$ & $0.0050$ & $0.0007$ & $0.0032$& $0.0027$ \\
$\sigma_{\Delta S}$ (kJ/kg/K) & $0.0077$ &  $0.0087$ & $0.0122$ & $0.0052$ & $0.0040$& $0.0224$ & $0.0113$ \\
$\sigma_{\Delta \mu}$ (kJ/kg) & $0.1582$ &  $0.3744$ & $0.1656$ & $0.1009$ & $0.0227$& $0.0746$ & $0.0836$\\
\hline
\end{tabular*}
\end{table}

The next point we address is to assess the ability of the models to predict accurately the properties for compressed liquids. We show in Fig.~\ref{Fig6} the variations of the density of water with pressure at ambient temperature ($300$~K). We also provide in Table~\ref{Tab4} the standard deviations obtained for each model with respect to the experiment. Remarkably, we find that the order in which the predicted densities appear on Fig.~\ref{Fig6} differs from the order obtained for the results at coexistence. The TIP4P/Ew model gives the most accurate account of both the absolute value for the density and its dependence on pressure up to $500$~bar, with the OPC and TIP5P models being close second and third, as confirmed by the standard deviations of Table~\ref{Tab4}. We add that the TIP4P/2005, which performed quite well for the VLE, slightly overestimate the liquid density and is ranked fourth in terms of accuracy for this property. This ranking is consistent with the predictions obtained at atmospheric pressure, which show that both the TIP4P/Ew~\cite{horn2004development} and OPC~\cite{izadi2014building} models yield densities that are within $0.002$~g/cm$^3$ of the experiment. Overall, we add that the TIP4P/Ew model performs particularly well for the density of liquid water under ambient pressure, since it also accurately predicts, within 1~K, the temperature of maximum density, while, on the other hand, the SPC/E, TIP3P and TIP4P are unable to predict a temperature of maximum density within 25~K of the experimental data~\cite{vega2011simulating}.

\begin{figure}
\begin{center}
\includegraphics*[width=6.5cm]{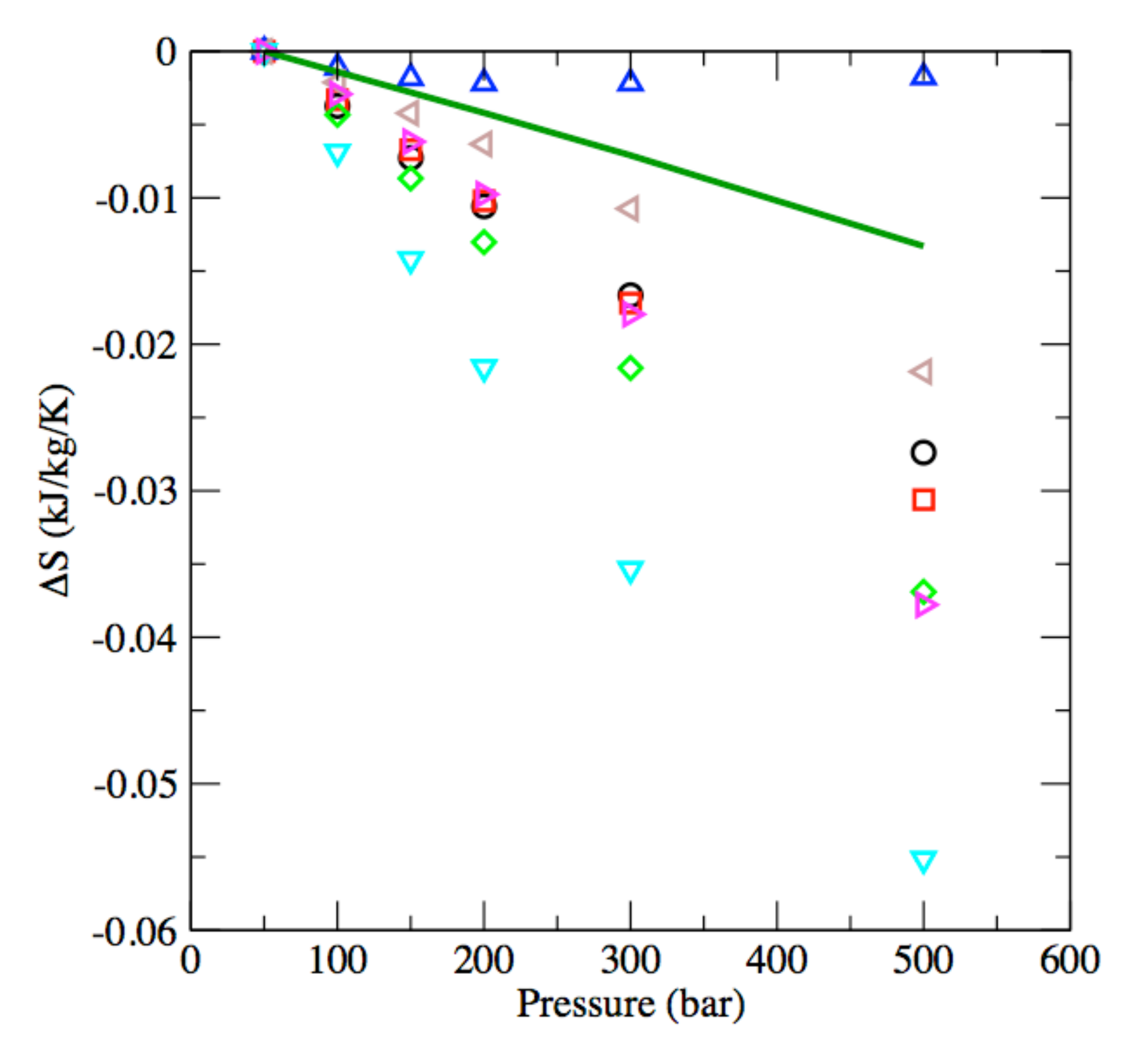}(a)
\includegraphics*[width=6.3cm]{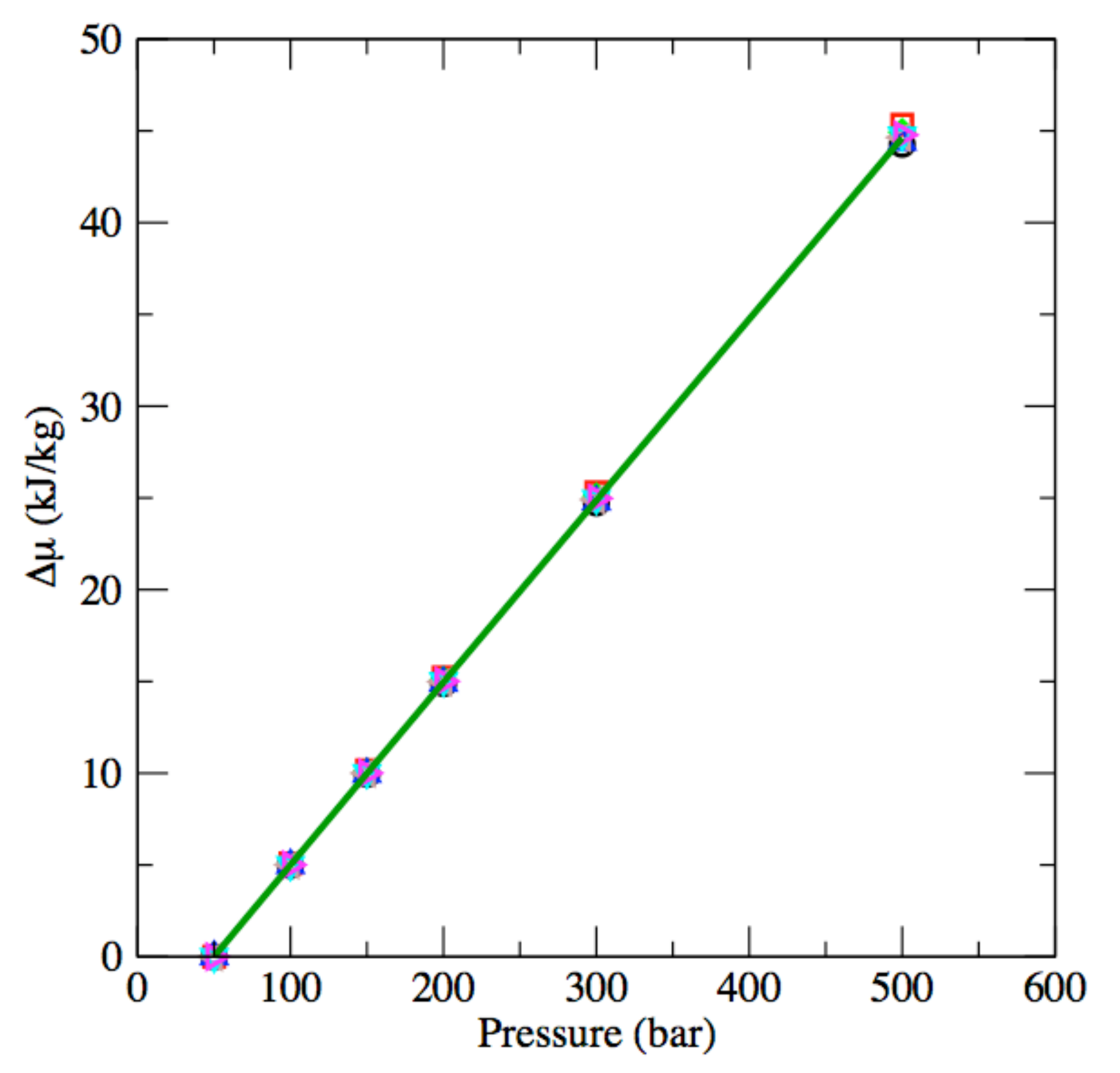}(b)
\end{center}
\caption{(a) Entropy change in compressed water at $T=300$~K (same legend as in Fig~\ref{Fig5}). Error bars are smaller than the symbol size. The entropy at $P=50$~bar is taken as the reference for each system. (b) Change in $\mu$ of compressed water at $T=300$~K (same legend as in Fig~\ref{Fig5}). Error bars are smaller than the symbol size. $\mu$ at $P=50$~bar is taken as the reference for each system.}
\label{Fig7}
\end{figure}

How does this impact the entropy and the molar Gibbs free energy of the compressed liquid? To carry out this analysis, we define as the reference value the entropy and molar Gibbs free energy at $P=50$~bar for each set of data. This allows us to plot the variations of these two functions against pressure in Fig.~\ref{Fig7}(a) and Fig.~\ref{Fig7}(b). We see in Fig.~\ref{Fig7}(a) that the reference data~\cite{wagner2002iapws} for the entropy change of compressed water falls between the predictions from the TIP4P/2005 and the TIP4P/Ew force fields, the latter being slightly more accurate than the TIP4P/2005 model in terms of standard deviations (see Table~\ref{Tab4}). Fig~\ref{Fig7}(a) highlights that a very accurate prediction of a liquid density does not automatically yield a similar accuracy for the entropy. For instance, the TIP5P model, which preformed very well for the density-pressure plot, gives the least accurate result for the entropy. This is unlike what we observed for $\Delta S_{vap}$. In that case, the dominant factor is the entropy of the vapor, which is closely related to the vapor density, since there is little organization in a dilute phase. Hence, the ability of a specific force field to predict correctly the vapor density will directly correlate with its ability to predict accurately $\Delta S_{vap}$. On the other hand, a condensed phase, like a compressed liquid, undergoes very moderate density changes and the amount of structural order is expected to have a significant effect on the entropy of a condensed phase. This implies that accurate density predictions for a specific force field do not translate directly into accurate predictions for the entropy of a condensed phase. As a result, despite the lower rankings obtained by the SPC/E and TIP3P models for the density-pressure results, these two models rank third and fourth for the entropy predictions, as shown by the standard deviations of Table~\ref{Tab4}. We perform the same analysis for the variations in the molar Gibbs free energy of compressed water as a function of pressure. Overall, all force fields provide a reasonable account of the dependence of the molar Gibbs free energy on $P$. Fig~\ref{Fig7}(b) shows a plot of $\Delta \mu$ against $P$, which allows for a comparison between the experiment and the $7$ force fields studied in this work. This graph confirms that the TIP4P/Ew force field models accurately the $\Delta \mu$-$P$ relation for compressed water. The next three force fields are ranked as follows, with TIP5P coming in third, OPC in fourth and TIP4P/2005 in fifth as shown in Table~\ref{Tab4} through the standard deviations from the experiment. Considering the results for the three properties ($\rho$, $\Delta S$, $\Delta \mu$) for compressed water, TIP4P/Ew appears to be the best rigid model, among those studied in this work, for molecular simulation studies of compressed water. 

\section{Conclusions}
In this work, we perform EWL simulations on a series of 7 rigid models for water, the SPC/E, TIP3P, TIP4P, TIP4P/2005, TIP4P/Ew, TIP5P and OPC force fields. The EWL method combines the advantages of a highly efficient flat histogram sampling, the Wang-Landau scheme, and of the expanded ensemble method to sample extensively the grand-canonical ensemble. It allows us to obtain a highly accurate value for the grand-canonical partition function of water over a wide range of temperature, from ambient temperature up to $475$~K. Using the statistical mechanics formalism, we compute the thermodynamic properties of water at the vapor-liquid coexistence and for compressed water. We validate our approach by comparing our results to those obtained using another flat histogram method, the TMMC method~\cite{errington2003direct,NIST}, and assess the ability of the force fields to predict accurately the properties of water through comparison to the experimental data. The results for the vapor-liquid equilibria establish the close connection that exists between the partition functions and the onset of the vapor-liquid transition, leading us to gather the behavior of the force fields into 2 subgroups, the \{TIP3P, TIP4P \} and \{SPC/E, TIP4P/2005, TIP4P/Ew\} subgroups, bracketed by the TIP5P and OPC force fields. Comparing the results for the equilibrium densities and the entropy of vaporization to the available data, we find that the predictions from the \{SPC/E, TIP4P/2005, TIP4P/Ew\} subgroup are the closest to the experiment for the VLE. The results also enable us to shed light on the existence of compensation effects, since the \{TIP3P, TIP4P \}  subgroup provides a reasonable account of the variations of $\Delta S_{vap}$ with $T$ despite their significant underestimation of the liquid density at coexistence. Looking at the properties for compressed water, we find that, of the $7$ models, the TIP4P/Ew model provides the closest match to the experimental data, both in terms of the variations of density, entropy and molar Gibbs free energy with pressure. In addition to providing benchmark entropies and free energies, as well as a ranking of the ability of the $7$ rigid force fields to model the thermodynamic properties of water, our results suggest that statistical (partition) functions could be used to further develop and parametrize water models.

{\bf Acknowledgements}
Partial funding for this research was provided by NSF through CAREER award DMR-1052808.\\

\bibliography{Water}

\break

\break

\section{Graphical Abstract}

\begin{scheme}
\includegraphics*[width=7cm]{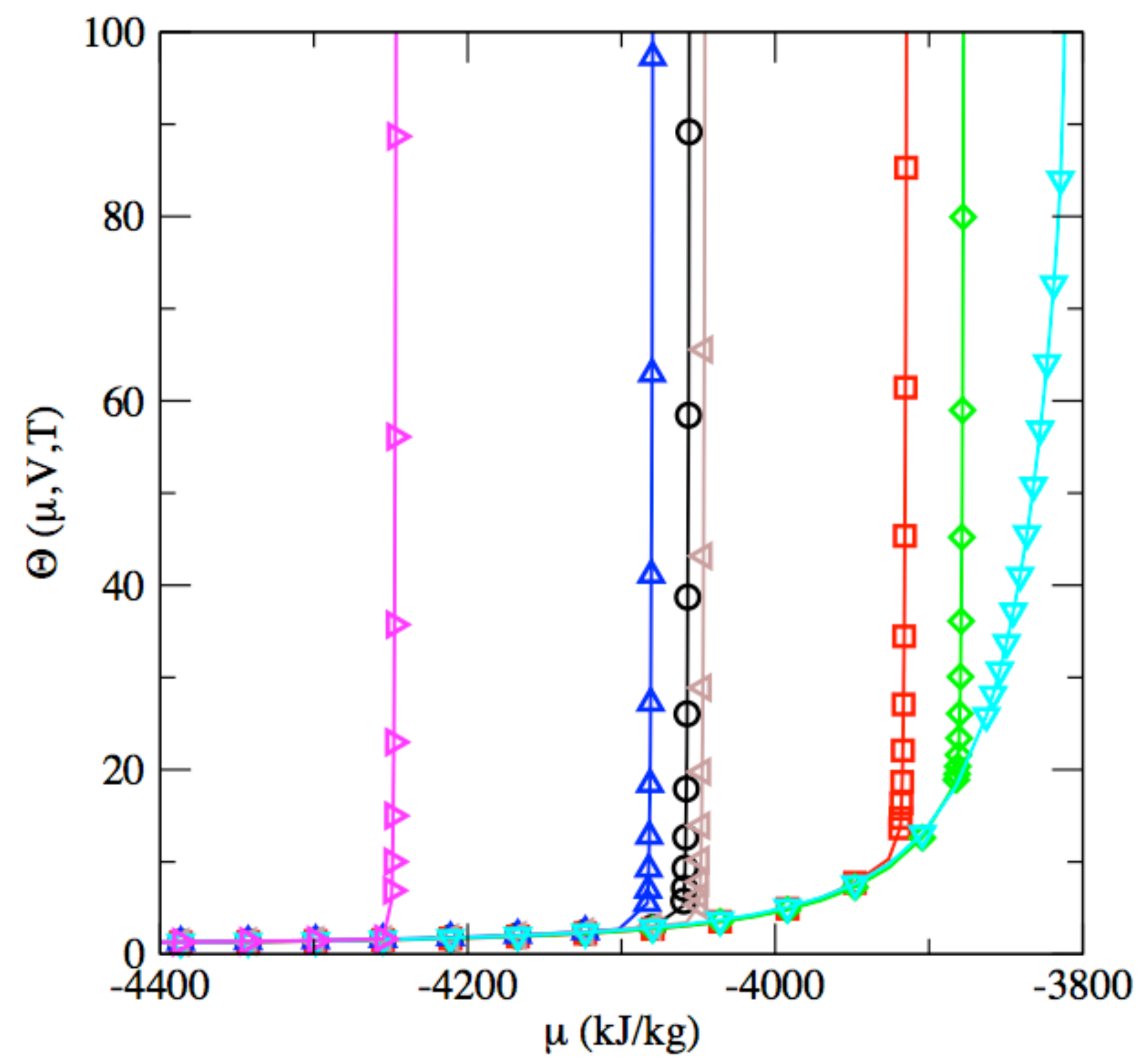}
\end{scheme}

\end{document}